\documentclass{aa}  
\usepackage{graphicx}
\usepackage{txfonts}
\def\secondip{\hbox{\rlap{\hbox{.}}\hbox{$''$}}}

\def\gradip{\hbox{\rlap{\hbox{.}}\raise 5.truept \hbox{{\small $\circ$}}}}

\begin{document}
\input{psfig}
%
% Next 5 lines define \simless and \simgreat: "less than or approximately
% equal to" and "greater than or approximately equal to".
\newbox\grsign \setbox\grsign=\hbox{$>$} \newdimen\grdimen \grdimen=\ht\grsign
\newbox\simlessbox \newbox\simgreatbox
\setbox\simgreatbox=\hbox{\raise.5ex\hbox{$>$}\llap
     {\lower.5ex\hbox{$\sim$}}}\ht1=\grdimen\dp1=0pt
\setbox\simlessbox=\hbox{\raise.5ex\hbox{$<$}\llap
     {\lower.5ex\hbox{$\sim$}}}\ht2=\grdimen\dp2=0pt
\def\simgreat{\mathrel{\copy\simgreatbox}}
\def\simless{\mathrel{\copy\simlessbox}}
% Next lines define "approximately proportional to"
\newbox\simppropto
\setbox\simppropto=\hbox{\raise.5ex\hbox{$\sim$}\llap
     {\lower.5ex\hbox{$\propto$}}}\ht2=\grdimen\dp2=0pt
\def\simpropto{\mathrel{\copy\simppropto}}

\title{Distances of the bulge globular clusters  Terzan 5, Liller 1, UKS 1 
and Terzan 4 based on HST NICMOS photometry}

%\subtitle{}

\author{
S.Ortolani\inst{1}
\and
B. Barbuy\inst{2}
\and
E. Bica\inst{3}
\and
M. Zoccali\inst{4}
\and
A. Renzini\inst{5}
\fnmsep
\thanks{Based on observations collected with the NASA/ESA {\it Hubble Space
Telescope} obtained at the Space Telescope Science Institute, which is
operated by the Association of Universities for Research in Astronomy,
Inc., under NASA contract NAS 5-16555.} }
\offprints{B. Barbuy}
\institute{
Universit\`a di Padova, Dipartimento di Astronomia, Vicolo
dell'Osservatorio 2, I-35122 Padova, Italy;\\
e-mail: sergio.ortolani@unipd.it
\and
Universidade de S\~ao Paulo, IAG, Rua do Mat\~ao 1226,
Cidade Universit\'aria, S\~ao Paulo 05508-900, Brazil;\\
e-mail: barbuy@astro.iag.usp.br
\and
Universidade Federal do Rio Grande do Sul, Departamento de Astronomia,
CP 15051, Porto Alegre 91501-970, Brazil;\\ e-mail: bica@if.ufrgs.br
\and
Universidad Catolica de Chile, Department of Astronomy \& Astrophysics,
Casilla 306, Santiago 22, Chile;\\ e-mail: mzoccali@astro.puc.cl
\and
Osservatorio Astronomico di Padova, 
Vicolo dell'Osservatorio 5, I-35122 Padova, 
Italy;\\  e-mail: alvio.renzini@oapd.inaf.it
}

\date{Received; accepted }

\abstract
{A large number of pulsars and X-rays sources are detected in globular clusters. 
To understand the structure and content of these clusters, accurate 
distances are required. }
{We derive the distances of  Terzan 5, Liller 1 and UKS 1  using as 
a reference a recent  distance determination  of NGC 6528, based on HST/NICMOS and 
NTT/SOFI infrared photometry. The distance of the metal-poor cluster Terzan 4
was derived from a comparison with M92 in NICMOS bands. }
%methods heading (mandatory)
{Distances of the metal-rich clusters are obtained by comparison 
of the Horizontal Branch  (HB) level of the clusters, relative 
to the reddening line passing through the HB of NGC~6528.
We use methods based on NICMOS bands and transformations to J and H 
magnitudes with different assumptions.}
{Liller~1 and Terzan~4 are found to be at the central bulge distance, UKS~1 is 
beyond the Galactic center, while Terzan 5 is closer to the Sun than the
other four clusters.}
{The distance of Terzan 5 is of paramount importance, given the impact of its
population of 21 pulsars, which is related to the high cluster density. 
The distance of Terzan 5 is  found to be d$_{\odot}$=5.5$\pm$0.9 kpc from the Sun, 
thus  closer to us than values given in studies of pulsars in Terzan 5. As a 
consequence, the higher cluster density is even more favourable for formation of 
the millisecond pulsars  recently detected in this cluster.}

\titlerunning{Distances of Terzan 5, Liller 1, UKS 1 and Terzan 4}
\maketitle

%
%________________________________________________________________

\section{Introduction}

Globular  cluster  distances  are  fundamental   to studies of  the  Galaxy
structure and stellar population  components
(e.g. Bica et al. 2006). Distances of  clusters near the Galactic center
are  particularly interesting, since several    of them host a   large
number  of X-ray sources  and pulsars (Cackett et  al. 2006; Ransom et
al.  2005; Lorimer 2005). 
Accurate distances are  crucial to estimate cluster stellar densities,
which in turn affect  stellar interaction rates in  globular clusters,
where dense environments favour formation of  binaries. Verbunt \& Hut
(1987; see also Hut et al. 1992) reported that 20\% of low-mass X-ray
binaries are  found in globular clusters.  They  also pointed out that
Terzan  5   and  Liller  1 might   be  the   densest   clusters in the
Galaxy. Ransom et al.  (2005) and Lorimer (2005) reported 
21 pulsars in Terzan 5.  Camilo \& Rasio (2005) reported that Terzan 5
and 47 Tucanae together have 45 millisecond  pulsars,  as a probable
result of the ``recycling'' model (where an  accreting neutron star is
spun up and becomes a fast radio pulsar). This model connects low mass
X-ray binaries (LMXB) to millisecond pulsars (MSP).

 The metal-rich globular cluster NGC 6528  is a template object of the 
bulge population (Bica 1988; Ortolani et al. 1992; Ortolani et al. 1995), 
and is used here as a reference. The distance of NGC 6528 was derived
by   Momany   et al. (2003) from
a combination of proper motion cleaned optical and infrared 
Colour-Magnitude Diagrams (CMDs),  obtained   from images observed with 
HST-WFPC2, and  the  ESO NTT-SOFI infrared camera.
 The resulting CMDs were compared to fiducial lines of  47 
Tucanae and NGC 6553 and to Padova isochrones.
A   distance modulus of (m-M)$_{\circ}$=14.44$\pm$0.06 
and a distance from  the  Sun d$_{\odot}$=7.7$\pm$0.2 kpc
were obtained,  where the errors were derived
from a comparison between  empirical and a theoretical calibration
of the HB. The distance errors in the infrared are small
relative to the optical, because of the minimized reddening dependence.
Momany et al.'s cleaned CMDs show unprecedented accuracy 
at the HB level of K(HB) =  13.20$\pm$0.05.  
A re-check taking into account calibration uncertainties,
indicates that the distance modulus error 
should be of the order of $\pm$0.10. Note that
the distance of the Galactic center considered previously to be
at 8 kpc (Reid 1993) has been recently revised to be closer
to the Sun at 7.62$\pm$0.32 kpc (Eisenhauer et al. 2005),
    7.5$\pm$0.1 kpc (Nishiyama et al. 2006) and
7.2$\pm$0.3 (Bica et al. 2006).

Ortolani et al. (2001) studied HST-NICMOS Colour-Magnitude Diagrams of
the reddened bulge globular  clusters NGC  6528,  Terzan 5,  Liller 1,
Terzan 4 and UKS 1, to derive their ages. The 
results confirmed that these bulge clusters are old, with ages comparable
with that of 47~Tucanae, as previously found by Ortolani et al. (1995).
In the  present   work,  we  concentrate efforts  to  derive 
distances for this sample,  taking NGC 6528  as a reference.

From the same  NICMOS data, Cohn et  al. (2002) derived a distance  of
Terzan~5 from  the Sun of d$_{\odot}$=8.7  kpc.  From optical NTT-SUSI
photometry, Ortolani et al. (1996a) derived d$_{\odot}$=5.6 kpc. In the
present paper, we readdress the Terzan 5 distance issue, given its
impact  on the   cluster  density  implying in binary  and   pulsar formation.
Matsunaga  et  al.  (2005)    searched  for  SiO masers   in  globular
clusters and a very  luminous one was  found in  the  central part  of
Terzan 5,  proposed to result from
 the merging of two AGB  stars, likely to occur  at this
 extreme stellar density.
  Also, Cohn et al.  (2002) reported an RR Lyrae in
Terzan 5, the  first detected at such high  metallicities,  and it was
interpreted as due to interaction effects.

Distances based on optical CMDs were derived
for the bulge globular clusters Terzan 4 (Ortolani et al. 1997a),
Terzan 5 (Ortolani et al. 1996a),
Liller 1 (Ortolani et al. 1996b)
and UKS 1 (Ortolani et al. 1997b).
The purpose of  this  paper is to derive more  accurate distances for 
these clusters 
making use of NICMOS data, as well as  to
better understand the large number of pulsars  in Terzan 5.
Distances of central clusters are also important to establish a 
locus of survival of old star clusters  in the central regions
of the Galaxy.

The observations and reductions  are briefly reported in Sect. 2.
In Sect. 3 we discuss cluster metallicities, based on the Red
Giant Branch (RGB) slope in the infrared.
 In Sect.  4  calibrations of NICMOS m$_{\rm 110}$ and  m$_{\rm 160}$
photometry are discussed.
In Sect. 5, distances derived in the present work are presented.
 In Sect. 6 implications for
the pulsar rich cluster   Terzan 5 are developed.  Concluding  remarks
are given in Sect. 7. Details on calibrations of NICMOS photometry are
given in Appendix A.

%___ 2 ________________________________________________________________

\section{Observations and reductions}

Ortolani  et  al. (2001)  observed  the sample globular  clusters with
NICMOS on board  HST, through the F110W and  F160W filters, using NIC1
and NIC2 cameras. The present analysis is based mostly on the NIC2 frames.  
The  pixel
size  of    NIC2 is  $0\secondip075$,  giving    a  field of   view of
$19\secondip2\times19\secondip2$  for each  frame. The description
of data, reduction, and a 
Log of the observations  are detailed in  Ortolani et al. (2001).  The
typical FWHM of  the stars was $0\secondip11$.  The zero point PHOTZPT
was applied to transform instrumental magnitudes into the standard HST
 magnitudes m$_{110}$ and m$_{160}$.

All the CMDs reach well below the HB.  The CMDs of
NGC 6528 and Terzan 4 reach  more than 3  magnitudes below the  turnoff 
in both the NIC1  and NIC2-offset pointings.  
The  deepest  CMD for Terzan 5 reaches about 1 mag. below the turnoff.

\section {Metallicities of the sample clusters}

Metallicities of globular clusters can  be  estimated from  the  RGB slope
as previously applied to optical and infrared CMDs (Ortolani et al. 1990, 1991;
Tiede et al. 1997).

\begin{figure}
\psfig{figure=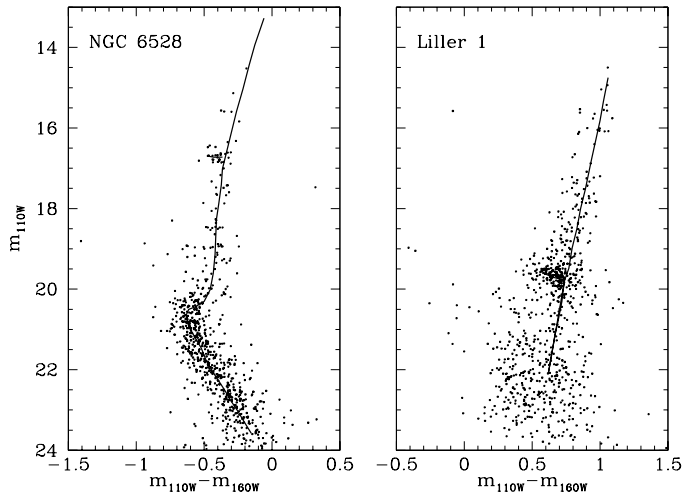,angle=-90,width=9.5cm}
\caption{ Colour-Magnitude Diagrams and mean loci 
in m$_{\rm 110}$ vs. m$_{\rm 110}$ --  m$_{\rm 160}$
for NGC 6528 and Liller 1.}
\label{fig1}
\end{figure}

\begin{figure}
\psfig{figure=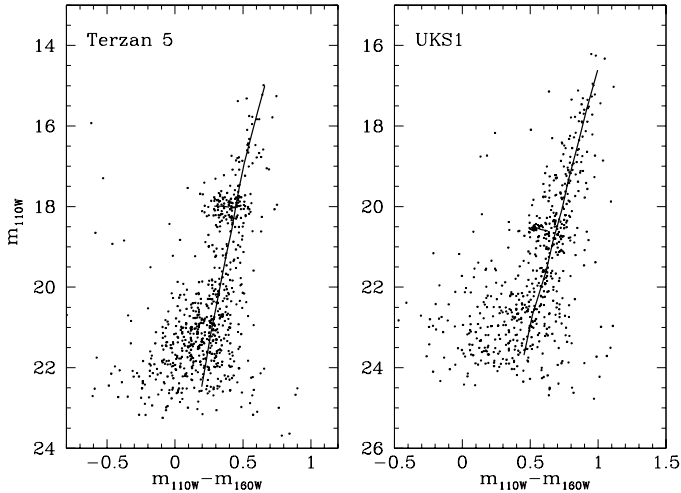,angle=-90,width=9.5cm}
\caption{ Colour-Magnitude Diagrams and mean loci in 
m$_{\rm 110}$ vs. m$_{\rm 110}$ --  m$_{\rm 160}$
for Terzan 5 and UKS 1.}
\label{fig2}
\end{figure}

 In Figs. \ref{fig1} and \ref{fig2}  we show the 
individual CMDs of the four metal-rich sample clusters,
and corresponding mean loci, plotted by eye.
We estimate an uncertainty of $\pm$0.1 
 in both  m$_{\rm 110}$ and  m$_{\rm 110}$ --  m$_{\rm 160}$.
In Fig. \ref{fig3} we also show the metal-poor cluster
Terzan 4 (Sect. 4.1).

 In Fig.~\ref{giants} are plotted
the mean loci of the sample clusters, by matching their HB level,
and subgiant branch (SGB) at the HB colour, or the turn-off in the case of 
Terzan 4. 
It appears that NGC 6528 is the most metal-rich cluster
of the sample, closely followed
by Terzan 5, and next by  Liller 1 and UKS 1, in this order. 
 The RGB of  Terzan 4 is
steeper due to its low metallicity.
The RGB slope measured in the CMD of NGC 6528 is 
$\Delta$m$_{\rm 110}$/$\Delta$(m$_{\rm 110}$ --  m$_{\rm 160}$) = 
11.0, while for UKS 1 it is  $\Delta$m$_{\rm 110}$/$\Delta$(m$_{110}$
--m$_{\rm 160}$)=15.0$\pm$1.0, the latter value being the most uncertain 
one.

Recent high resolution spectroscopic determinations confirm this
ranking for NGC 6528 (Zoccali et al. 2004), Terzan 5 and Liller 1
 (Origlia et al. 2002; Origlia \& Rich 2004), 
and UKS 1 (Origlia et al. 2005). 

Reddening and metallicity values for the sample clusters  
revised by Barbuy et al.  (1998) and the  compilation by Harris (1996),
as given at http://www.phy\-sics.mc\-mas\-ter\-.ca/Glo\-bu\-lar.html,
as well as recent spectroscopic determinations are shown in Table 1.

\section{Distance from NICMOS instrumental magnitudes}

For the derivation of distance, two methods are possible,
one using comparisons with template clusters in the instrumental
NICMOS bands, and the second by transforming them into J and H.

 Schlegel et  al. (1998)'s reddening law  (1998) was employed
to obtain the infrared extinction slope 
A$_{110}$/(A$_{\rm 110}$-A$_{160}$)  in  the     
HST/NICMOS bands:   A$_{110}$/E(B-V)=1.186,   
A$_{160}$/E(B-V)=0.634,   and     
A$_{110}$/(A$_{110}$-A$_{160}$) =  2.148. 
As a check, we also calculated the slope using the Lee et al. (2001)
 NICMOS extinction law based on synthetic spectra,
 obtaining a similar value
 A$_{110}$/(m$_{110}$-m$_{160}$) = 2.03.

Fig.~\ref{mloci} shows  the  observed mean loci  for  the  5  clusters,  in 
NICMOS  m$_{\rm 110}$ vs. m$_{\rm 110}$-m$_{\rm 160}$.
The  reddening line A$_{110}$/E(m$_{110}$-m$_{160}$) = 2.148
passing through the HB of NGC~6528 is also shown in Fig.~\ref{mloci}.
  Were all
the clusters at the same distance, then the HB of all of them would be
on  the same line.   Therefore, the m$_{110}$ difference between
each  cluster HB and  the  reddening   line represents  the  distance
modulus difference between NGC 6528 and the cluster, assuming that
they share the same
metallicity (or same intrinsic SGB colour and HB absolute value). 
The advantage of this method is that reddening from NICMOS
colours is not needed.
% The
%corresponding $\Delta$E(m$_{110}$-m$_{160}$)  is the resulting 
% reddening relative  to NGC 6528. 

Fig.~\ref{mloci} shows that Liller 1 and   NGC 6528 differ  
significantly  in reddening.
Terzan 5 is  located  above the reddening line,  thus  at a  shorter  distance,
whereas UKS 1 is below the line, and therefore farther than NGC 6528. 
The distances derived are given in column 5 of Table 2.

\subsection{Distance of the metal-poor cluster Terzan 4}

\begin{figure}
\psfig{figure=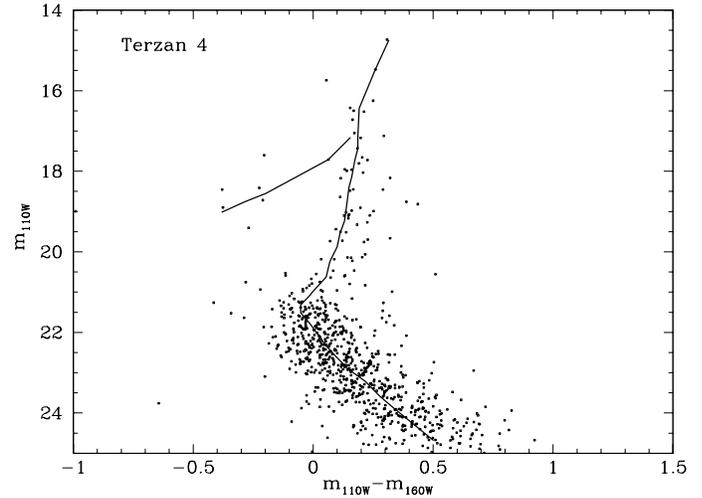,angle=-90,width=9.5cm}
\caption{Colour-Magnitude Diagram of Terzan 4,
 and mean loci of M92 overplotted, in 
m$_{\rm 110}$ vs. m$_{\rm 110}$ --  m$_{\rm 160}$.
  }
\label{fig3}
\end{figure}

\begin{figure}
\psfig{figure=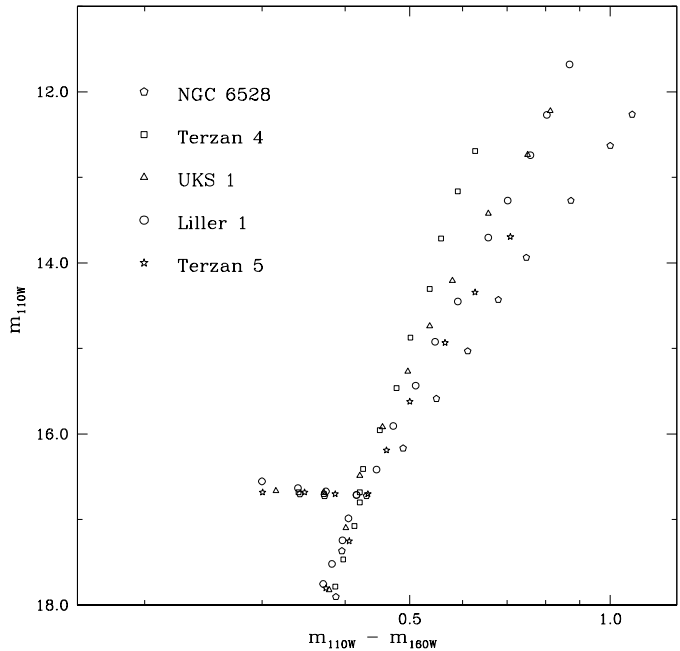,angle=0,width=9.5cm}
\caption{Mean loci in m$_{\rm 110}$ vs. m$_{\rm 110}$ --  m$_{\rm 160}$
for the five sample clusters, by matching relative to NGC 6528 their HBs
and SGB@HB, or the turn-off in the case of Terzan 4. 
NGC 6528: black circles
 Terzan 5: open triangles; UKS 1: black triangles;
 Liller 1: open pentagon; Terzan 4: black square.
  }
\label{giants}
\end{figure}

\begin{figure}
\psfig{figure=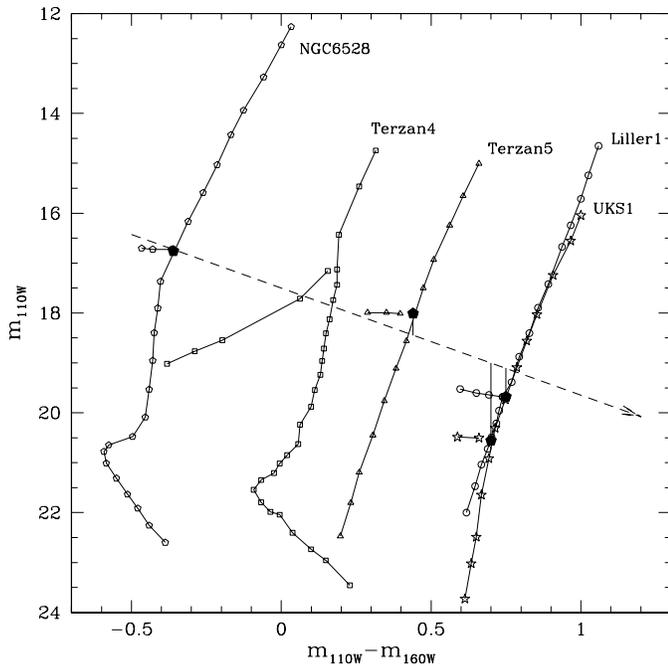,angle=0,width=9.5cm}
\caption{Mean loci in m$_{\rm 110}$ vs. m$_{\rm 110}$ --  m$_{\rm 160}$
for the five sample clusters, in ST magnitudes.  The solid line is the
reddening    line   with     a    slope    A$_{\rm    110}$/E(m$_{\rm
110-160}$)=2.15, passing through NGC 6528.
NGC 6528: open circles; Terzan 4: open squares;
 Terzan 5: open triangles; Liller 1: open circles; UKS 1: open stars.
  }
\label{mloci}
\end{figure}

Given the difference in Horizontal Branch morphology
between metal-poor and metal-rich clusters, the method described
in Sect. 4 cannot be applied to Terzan 4.
The distance of Terzan 4 was derived by comparing its
 CMD to the mean locus of M92 (Lee et al. 2001) observed in
the same NICMOS bands (Fig. \ref{fig3}).
 Lee et al.'s 
 reddening law in the NICMOS bands was adopted.
The best fit  gives
$\Delta$(m$_{\rm 110}$-m$_{\rm 160}$)(Tz4-M92) = 0.73$\pm$0.02 and
$\Delta$(m$_{\rm 110}$)(Tz4-M92) = 1.6$\pm$0.1. Adopting
E(B-V) = 1.87 E(m$_{\rm 110}$-m$_{\rm 160}$) = 1.38 
and A$_{\rm 110}$/E(B-V)=1.16 (Lee et al. 2001), we get
A$_{\rm 110}$ = 1.38 x 1.16 = 1.60 and
$\Delta$(m-M)$_{\circ}$(Tz4-M92) = 0.0.
Finally A$_{\rm V}$ = 1.38 x 3.3 = 4.5 mag.

This means that Terzan 4 is at the same distance as  M 92,
which is 8 kpc according to Harris (1996).  Therefore the distance of 
 Terzan 4  is R$_{\odot}$=8.0$\pm$0.3 kpc.
This is very close to the Galactic center. Its extinction is
rather below the 7 magnitudes we derived in the optical, 
but the distance is compatible with our previous estimations of
 8.3$\pm$0.7 kpc (Ortolani et al. 1997a) and 7.3 kpc  (Barbuy et al. 1998).
The present determination gives the most reliable
 distance and reddening values for Terzan 4.

\subsection{Errors}

Distance errors in the method described above 
 are dominated by  uncertainties on:

(a) The  measurements of SGB colours and  luminosities at the level of
the HB. Differential reddening across the images is the main source
of errors, and such effects are common to all calibrations.
Differential reddening together with some field contamination of the
lower Galactic latitude clusters (such as Liller 1) are the main causes
of scattering in the CMDs.
  The intrinsic  photometric errors are negligible  because
of the very high signal to noise ratio at the HB level, given that the
photometry  reaches below the  turnoff. In the
CMDs they are of the order of 0.07 in colour and 0.15 in magnitude.  
These values were obtained by   measuring the HB  in  different parts  of  the 
images
and/or comparing  NIC1 and NIC2  cameras. However, since the CMD
scatter is mainly  in the direction of  the reddening line, the effect
propagated on the distance is reduced to about 0.1 mag. A metallicity
uncertainty of about 0.2dex would imply  the same error (0.1 mag).

(b) The slope of the reddening law.  The formal error on the distances
introduced  by  the uncertainty  on the  slope of  the  reddening line
(Schlegel et al. 1998; Fitzpatrick 1999) is considerably lower, of the
order of 0.02 mag. in the full colour range.  Second order effects may
arise due to the wide passband of the F110W and F160W filters (the first
is  about twice that of the standard  J, and the  second 35\% wider than the H
band). Possible non linear effects due to  the distorted passbands may
be an  important uncertainty  of this calibration  and  cannot be
easily quantified. While the  second order reddening effects are  well
known in the optical   photometry,  in the infrared   the effects  on  the
standard K band have been only recently inferred (Kim et al. 2005).

\section{Reddening and distance from NICMOS calibrated data}

 J and H calibrated magnitudes and colours have  been obtained by adopting
linear transformation equations  by Stephens et al.  (2000) 
 and  Cohn et  al. (2002).  Both authors used the NICMOS  
Data Handbook standard stars,
whereas  Stephens et  al.  added Baade's  Window giants  observed both with
NICMOS and from ground in JHK. More details on the calibrations
are given in the Appendix.

In  Fig.~\ref{dist3} we show  the J  level of the HB
vs. $J-H$  of the SGB  at   the HB level   (SGB@HB) for  the 4 metal-rich
 sample clusters. In  Figs. \ref{mloci} and \ref{dist3}, the distance determination
is   independent of reddening  values  and zero  point calibration.  A
source of uncertainty is the  choice of the reddening  law, but 
differences  between them in the infrared are  small
(Schlegel et al. 1998; Fitzpatrick 1999). Moreover, 
there is a general agreement in the literature that, in the infrared,
 the reddening
slope is constant and independent of the dust characteristics.

Transformations of E(J-H) into E(B-V) are reported in the Appendix.
Assuming  for NGC 6528 the  reddening  E(B-V)=0.46 (Zoccali et
al.  2004), we get for Terzan 5, Liller 1 and UKS  1 the $\Delta$E(B-V)
 values.  In order to obtain the total reddening,
that of NGC 6528 itself has to be added, and the results are
given in Table 2 (columns 9 and 13 for the two calibrations respectively).
A good agreement is found with  optical CMDs  (Table 1) for
Terzan 5. For Liller 1 and UKS 1, Cohn et al.'s calibration
is more compatible with the literature.

In Table 2 are given  distance determinations for the sample clusters,
assuming for NGC 6528 the value d$_{\odot}$=7.7 kpc from Momany et al.
(2003), and using  the independent  calibrations
from instrumental to J and H bands by
Stephens et al. (2000) and Cohn et al. (2002). 
Columns 8 and 12 of Table 2  give the distance modulus difference
between each cluster  and NGC 6528, measured along the reddening line
(Figs. \ref{mloci} and \ref{dist3}), for each calibration.  
This  makes  this  procedure 
reddening independent.  The differences in distances between
the two calibrations arise from the   
different  calibration coefficients,   and not  from 
photometric zero points.

In Table 2 we calculated the distances of UKS 1 and Liller 1 for two
metallicity values: the same as that of NGC 6528 ([Fe/H]=-0.1),
and a  correction of $\Delta$[Fe/H]=0.5,  making use of Padova isochrones
(Girardi et al. 2000).
The colour changes by
$\Delta$(J-H)$\approx$0.1, corresponding to a decrease of
 0.25 magnitudes in the distance modulus. 

Terzan 5 distance estimates of 4.6 and 5.6 
kpc, put it closer than the Galactic center, whereas
a mean of   6.5 and 9.1 kpc for Liller 1
place the cluster near the Galactic center.
UKS 1  estimates of  9.3 to 12.9 kpc suggest that the cluster
is beyond  the  Galactic center.

\begin{table*}
\begin{center}
\caption[1]{Literature reddening and metallicity 
(Barbuy et al. 1998; Harris 1996 and high resolution
spectroscopy by 1 Origlia \& Rich 2004; 
2 Carretta et al. 2001; 3 Zoccali et al. 2004; 4 Origlia et al. 2005; 5 Origlia et al. 2002)
}
\begin{tabular}{lllllllllll}
\hline
\noalign{\smallskip}
 &  &  & \multispan2 Barbuy et al. 1998 & &\multispan2 Harris' compilation \\
\cline{4-5} \cline{7-8} 
Cluster & l & b & E(B-V) & [Fe/H] &
& E(B-V) & [Fe/H] &  [Fe/H]spec & ref.  \\
\noalign{\smallskip}
\hline
\noalign{\smallskip}
NGC~6528  & $1\gradip14$   & $-4\gradip18$&0.52  &  -0.2 & &0.56 & -0.17  & -0.1 & 2,3 \\
Terzan 5  & $3\gradip81$ & $1\gradip67$ & 2.39  &  0.0 & & 2.37 & -0.28 & -0.21 & 1 \\
Liller 1  & $354\gradip81$ & $-0\gradip16$ &3.0  &  +0.2 &  & 3.00 & +0.22   & -0.3 & 5 \\
UKS 1     & $5\gradip13$ & $0\gradip76$  &3.1  & -0.5  & & 3.09  & -0.5  & -0.78 & 4 \\
Terzan 4  & $356\gradip02$ & $1\gradip31$  &2.31  & -2.0:&  & 2.35 & -1.60  & -1.6 & 1 \\
\noalign{\smallskip} \hline \end{tabular}
\end{center} 
\end{table*}

\section{Pulsars and density of Terzan 5}

  Terzan 5 distance estimates  range from 4.6 to 6.3
kpc (Table 2), shorter than Cohn et al. (2002)'s 
value of 8.7 kpc. 
Cohn et al. derived this distance in two ways:
(i) One using a comparison of HB level of NGC 6528 in the H band,
adopting H(HB)=13.1 from Davidge (2000). However it can be
 checked in Davidge's Fig. 5 that instead H(HB)=13.5 for NGC 6528,
and the same value is found in Momany et al. (2003). Terzan 5
H(HB) is also different: 
Cohn et al. adopt H(HB)=13.0 (in fact incompatible with
 his Fig. 6 that indicates H(HB)=14.0) whereas
we have H(HB)=13.76 and 13.9 according to Stephens et al. or Cohn et al.'s
calibrations.
(ii) Cohn et al.  used the Baade's Window Red  Giant  Branch as reference,
and deduced E(B-V)=2.18 for Terzan 5. Using the distance from
Ortolani et al. (2001) of 5.6 kpc, together with their lower 
reddening, a distance of 8.7 kpc was found for Terzan 5 by Cohn et al.

%  whereas   we use a   specific CMD
%location, which is the magnitude and colour of the SGB at the level of
%the HB relative to NGC 6528.
In conclusion, the incorrect H(HB) values in (i), and a low reddening
of (E(B-V)=2.18 for Terzan 5 in (ii), may explain 
Cohn et al.'s different distance estimates relative to ours.

In the present work (Table 2),
estimated distances of Terzan 5 are 6.3, 4.6 and 5.6 kpc. 
 The  three values are consistent, reinforcing the results, given
that the first method is independent of photometric calibrations.
An average distance d$_{\odot}$  =  5.5$\pm$0.9kpc is adopted  for
Terzan 5.

Camilo \& Rasio (2005)  presented evidence   that  globular clusters  with
pulsars have  densities  $\rho >$  1000 L$_{\odot}$/pc$^3$.  In  their
Fig. 6 Terzan 5 is  reported to be one  of the densest clusters,  with
$\rho >$ 10$^{5.5}$ L$_{\odot}$/pc$^3$, assuming however a distance of
7.6 kpc from the Sun (Harris 1996).
 
A  closer   distance  causes  a  higher   calculated   stellar density
(L$_{\odot}$/pc$^3$).  Luminosity  scales   with  R$^2$, whereas   the
volume scales with R$^3$, therefore the density increases for shorter
distances.  The present results imply that  the density would increase
by  30-40\%.

% $\rho \approx$ 10$^{5.6}$ L$_{\odot}$/pc$^3$.
% Terzan 5 would become then the densest cluster known.

%\begin{table*}
%\begin{center}
%\caption[1]{Observed $J-H$ and $J$ of the SGB at the level
%of the HB. The second line values are the distances resulting 
%by assuming the NGC 6528 distance as zero point. 
%The assumed reddening line is AJ/E(J-H)=2.76.
%a: value adopted from Momany et al. (2003).
%}
%\begin{tabular}{llllllllll}
%\hline
%\noalign{\smallskip}
%cluster& NICMOS$_{J-H}$ & NICMOS$_J$ & NICMOS$_J$$_{\rm calc}$  & $\delta$J & (m-M)$_{\circ}$ & E(J-H) & d$_{\odot}$(kpc) &    \\
%\noalign{\smallskip}
%\hline
%\noalign{\smallskip}
%NGC 6528 &    -0.36 &      16.71  &  & 0    &       14.44$^a$ &   0 & 7.7$^a$  \\
%Terzan 4 & obs  0.22  & 17.96  & 18.31 &  0.35  &      14.09*   &  0.57*   &   6.6* \\
%Terzan 5 & obs  0.45  &  18.0 & 18.97  &  0.97  &      13.47    &     0.80 &  4.9 \\
%UKS 1 & obs 0.70  & 20.56 &  19.67 &  -0.9  &      15.37    &     1.06   &  16.7  \\
%Liller 1 & obs 0.75  & 19.67 & 19.78 &   0.12  &      14.32     &   1.1 &  7.3   \\
%\noalign{\smallskip} \hline \end{tabular}
%\end{center} 
%\end{table*}

\begin{figure}
\psfig{figure=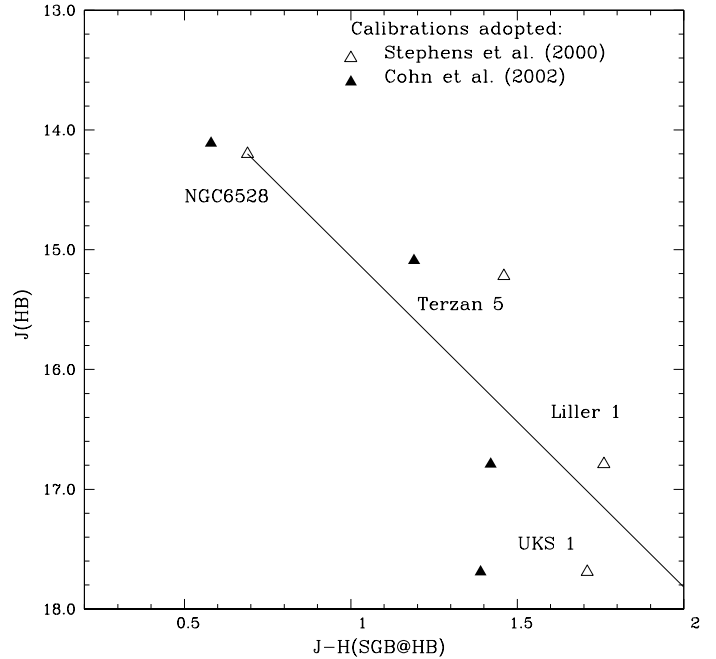,angle=0,width=9.5cm}
\caption{$J$ level of the HB vs. $J-H$ of the SGB at the HB level
for the four metal-rich clusters, derived with two different
transformation equations. The solid line is the reddening line
with a slope A$_{\rm J}$/E(J-H)=2.76, passing through NGC 6528.
Symbols: open triangles: transformation equations by Stephens et al.
(2000); filled triangles: transformation equations by Cohn et al. (2002).
}
\label{dist3}
\end{figure}

%===============================================================================
\begin{table*}
\begin{flushleft}
\caption{Horizontal branch magnitudes, colours and distances for 
different photometric calibrations. The
m$_{\rm 110}$ and J magnitudes refer to the HB, colours refer to the SGB at the HB level.
 $\Delta$m$_{\rm 110}$
and $\Delta$J are the distance moduli differences 
relative to the reference cluster NGC 6528.  Assuming for UKS 1 and Liller 1
 a  metallicity of $^a$ the same as NGC 6528 of [Fe/H]=-0.1; $^b$ more probable
value of [Fe/H]=-0.6.
}             
\label{calibs}      
\centering          
\begin{tabular}{l@{}cccccccccccccccc}     % 12 columns 
\noalign{\smallskip}
\hline\hline    
\noalign{\smallskip}
\noalign{\vskip 0.1cm} 
 & \multicolumn{4}{c}{NICMOS instrumental} &  \multicolumn{5}{c}{Stephens et al. calibration} & & \multicolumn{4}{c}{Cohn et al. calibration} &  \\
\cline{2-5} \cline{6-10} \cline{11-14}   \\
\hbox{Cluster} & \hbox{m$_{\rm 110}$-m$_{\rm 160}$} & \hbox{m$_{\rm 110}$} & \hbox{$\Delta$ m$_{\rm 110}$} & 
 \hbox{d(kpc)} & \hbox{J(HB)}  & \hbox{J-H(HB)}  & \hbox{$\Delta$J} & \hbox{E(B-V)} & \hbox{d(kpc)}
         & \hbox{J(HB)}   & \hbox{$\Delta$J} & \hbox{E(B-V)} & \hbox{d(kpc)} \\
\noalign{\smallskip}
\noalign{\vskip 0.1cm}
\noalign{\hrule\vskip 0.1cm}   
\hbox{NGC 6528} & -0.36 & 16.71 & ----   & 7.7  & 14.2 &0.58& ----  & 0.46  & 7.7      & 14.11     &  ----&0.46 &7.7 &  \\
\hbox{Terzan 5} &  0.44 & 17.99 & -0.44  & 6.3  & 15.22&1.19& -1.10 & 2.82 & 4.6 & 15.09 & -0.70 &2.33 &5.6 &  \\
\hbox{Liller 1$^a$} &  0.75 & 16.66 &  0.57  & 10.0 & 16.79&1.42& -0.36 & 3.74 &  6.5 & 16.79&  0.36 & 3.04 & 9.1 &  \\
\hbox{Liller  1$^b$} &    &       &        & 8.9 &      &    &       &      & 5.8 &      &       &  & 8.1 &   \\
\hbox{UKS 1$^a$}    &  0.70 & 20.56 &  1.57  & 15.9 & 17.69&1.39&  0.68 & 3.59 & 10.6 & 17.69&  1.34 &2.94 & 14.3 &   \\
\hbox{UKS 1$^b$} &    &       &        & 14.2 &      &    &       &      & 9.3  &      &       &  & 12.9 &   \\
\noalign{\vskip 0.1cm}
\noalign{\hrule\vskip 0.1cm}   
\hline                 
 \end{tabular}
\end{flushleft}
\end{table*}
%=====================================================================
\section{Conclusions}

Using  NICMOS on board   the    Hubble Space Telescope 
(Ortolani et  al.  2001),  we derive relative  distances for
 Terzan 5, UKS 1 and Liller 1, based on the distance value of
NGC 6528 derived by Momany et al. (2003). The distance of the metal-poor
cluster Terzan 4 was derived from a comparison with M92 in the same
NICMOS passbands.

From the diagram $J$(HB) vs. $J-H$(SGB@HB) (Fig.~\ref{dist3}), 
 Terzan 5 is found to  be closer to the Sun  than NGC 6528,
whereas Liller 1 is at a comparable distance, and UKS  1 appears to be
  more distant.    Assuming  a distance  to the  Galactic
center of R$_{\rm GC}$=8 kpc (Reid 1993), or closer distances of
R$_{\rm GC}$=7.2 to 7.6 kpc (Eisenhauer et al. 2005; Nishiyama et al. 2006; Bica et
al. 2006), NGC 6528 is essentially at the Galactic center distance.
Liller 1 could be somewhat beyond the Galactic center.

The relative metallicity of the sample metal-rich clusters, based on
RGB slope in the infrared, show that NGC 6528 is the most metal-rich
cluster, closely followed by Terzan 5, Liller 1 and UKS 1.
Terzan 4 is clearly more metal-poor. The lower metallicity of UKS 1
favours the shorter distance of 9.3 kpc based on
 Stephens et al. (2000)'s calibration, placing
it in the bulge volume.

We have shown that  Terzan 5 is  at d$_{\odot}$=5.5 kpc from  the Sun,
thus  significantly   closer than  the   value  of d$_{\odot}$=8.7 kpc
calculated  by Cohn et al.  (2002), or  d$_{\odot}$=7.6 kpc adopted by
Camilo  \& Rasio (2005).  
 Recently, Ransom (2006) pointed out that if the distance of Terzan 5
were really 8.7 kpc, then the current models of electron distribution
in the Galaxy would overestimate the integrated electron density towards
the cluster. For 8.7 kpc, models predict a dispersion measure of
530 pc/cm$^{3}$, while for the pulsars in Terzan 5, the dispersion
measured is estimated to be of 239 pc/cm$^{3}$. This discrepancy
is better explained by the  distance of 5.5 kpc.
 The present shorter  distance has also important
implications concerning the cluster  density.  A 30-40\% decrease   in
distance using  Camilo  \& Rasio's  (2005)  value leads to  a density
increase by the same fraction,  which is important for the calculation
of the stellar collision rate, leading to   the
formation of  low mass X-ray  binaries and millisecond pulsars in this
cluster (Edmonds et al. 2001; Ransom et al. 2005).

\appendix
\section{JH calibrations  of NICMOS photometry }

The  calibrations we analysed in detail are:

\noindent {\it a) Stephens et al.(2000): }

${\rm J = -0.344 (m_{\rm 110}-m_{\rm 160}) + const}$

${\rm J-H = 0.96 (m_{\rm 110}-m_{\rm 160}) + const}$

\noindent {\it b) Cohn et al. (2002):  }

${\rm J = -0.335 (m_{\rm 110}-m_{\rm 160}) + const}$

${\rm J-H = 0.68 (m_{\rm 110}-m_{\rm 160}) + const}$

\noindent {\it c) Girardi et al. (2000) for Z=0.02: }

${\rm J = -0.32 (m_{\rm 110}-m_{\rm 160}) + const}$

${\rm J-H = 0.77 (m_{\rm 110}-m_{\rm 160}) + const}$

\noindent {\it d) Girardi et al. (2000) for Z=0.0004:}

${\rm J = -0.2 (m_{\rm 110}-m_{\rm 160}) + const}$

${\rm J-H = 0.86 (m_{\rm 110}-m_{\rm 160}) + const}$

\subsection{Uncertainties in the colour transformation}

The errors on the distances and reddening values are dominated
by the transformation coefficients (colour terms), from the NICMOS
instrumental system, into the JH magnitudes. 
 More  reddened clusters are more dependent on
the reddening/calibration  slopes, which is   the case  of  UKS 1  and
Liller 1.

The calibration of wide band photometry in non-standard
systems, such     as  NICMOS, is  not  trivial.     The  estimation  of
uncertainties  has no simple solution because standard stars
are in general  not available  for  extreme temperature and  reddening
values.  Non-linearities  and   second order   effects may   be  non
negligible.     In the following  we    analyse  separately the  two
different calibrations that we used to derive  the distances
(Table 2), and compare them with other calibrations in the literature.

\subsection{Stephens et al.'s calibration}

Stephens et  al.'s calibration (equations a)
 has a  J  slope coefficient (-0.344) similar to
Cohn et al. (2002)'s calibrations,   but their  H slope is
larger (-0.305, vs.   $\approx$ -0.1). This provides  a  flat slope for
the J-H transformation equation,  and somewhat redder colours for very
red stars. This explains the relatively shorter distance scale for the
most reddened  clusters (Liller 1,  UKS 1), combined with 
high reddening. 

Stephens et al. (2000)  checked  fits with simulations  using infrared
spectra of reddened  stars.   Assuming that  bulge  field giants   are
similar to cluster  giants, this method potentially offers the  most
reliable transformations for bulge clusters. 
On the other hand, data for Baade's Window
stars  show some scatter,   and they  cover a  limited  colour  range,
producing errors of $\pm$0.1 on the colour slope and $\pm$0.08 mag. on
the zero point, and an error up to $\pm$0.2 mag at the extremes of the
colours of the sample clusters.

\subsection{Cohn et al.'s calibration}
The calibration used by Cohn et al. (2002) (equations b) is based
on the NICMOS five solar neighbourhood standard stars, distributed over  
 a wide temperature/colour  range.  The J-H colour term 
is lower than in Stephens et al.'s calibration (0.76 vs. 0.96), 
resulting in bluer colours and increasing distances 
for more  reddened clusters (Table 2). The
formal errors, from the linear  interpolation, are similar to those in
Stephens et al. They are
  marginally compatible with the high
metallicity of Liller 1 and UKS 1, and with Liller 1 JHK photometry 
(Frogel et al. 1995). The JH CMDs of NGC 6528 calibrated with
Stephens et al. and Cohn et al.'s equations have been compared
with Momany et al. (2003)'s ground based photometry. The colours
were found in better agreement
 with Stephens et al.'s calibration, whereas Cohn et al.'s
calibration gave somewhat bluer colours.

\begin{figure}
\psfig{figure=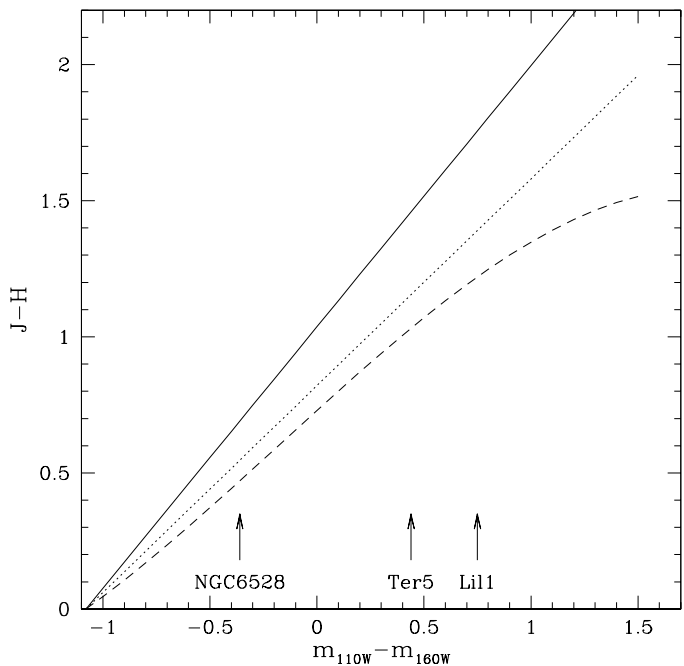,angle=0,width=9.5cm}
\caption{Colour calibrations from Stephens et al. (2000) (solid line), 
Cohn et al. (2002) (dotted) and Origlia \& Leitherer (2000)
(dashed). The m$_{\rm 110}$-m$_{\rm 160}$ colours of NGC 6528, 
Terzan 5 and Liller 1 
are shown.
}
\label{origlia}
\end{figure}

\subsection{Other calibrations}

  Cohn et al.'s calibration equations are similar
to others  in the literature, such as  Macri et  al.'s (2001).  In the
latter case the calibration was derived  from solar metallicity Kurucz
models, convolved with NICMOS  passbands and  compared to the
standard   JH system.   Reddening effects   were  not considered. They
derived small formal errors, one order of magnitude less than those of
Cohn et al. (2002).

A  colour calibration was obtained by Origlia \& Leitherer (2000).
They used Bessell et al. (1988) model atmospheres to transform NICMOS
colours into the JHK system, and concluded that the transformations
are gravity and metallicity independent, but a cubic polynomial relation
is required to fit the data. The effect of such an interpolation is that for
NGC 6528 and Terzan 5, with  reddening values lower than UKS 1 and Liller 1,
 the results
are similar to Cohn et al. (2002) and Macri et al. (2001) calibrations, but
the colours are considerably bluer for more reddened clusters, implying
very large distances for UKS 1 and Liller 1. The effect of 
linearity deviations is seen in their Fig. 11,
as well as a  discrepancy with the reddest standard stars. Fig. A.1 shows
the different colour calibrations in the metallicity range of our clusters.

From  the  analyses  above we  conclude  that 
systematic errors of the order of $\pm$0.2-0.3 mag., related mostly to
colour transformation  slopes, affect distance moduli, in particular those
of the most reddened clusters.

\subsection{Calibrations with Padova stellar evolution models}

In order to clarify the nature of the discrepancies between
 the described calibrations, we performed a further 
independent test based on the Girardi et al. (2002) NICMOS isochrones 
in the m$_{\rm 110}$ and m$_{\rm 160}$ bands. 
The authors derived the isochrones 
from ATLAS9 models  (Castelli et al. 1997) and a NICMOS transmission
system provided by the NICMOS staff. Comparing these isochrones with those 
by Girardi et al. (2000) in the JHK system, 
we obtained the transformation equations c) and d) for metallicities
 of Z=0.019  (solar), and a low
 metallicity of Z=0.0004, in the full colour range of the isochrones 
(about  0.2$<$m$_{\rm 110}$-m$_{\rm 160}$$<$1.1). For solar metallicity the
linear interpolation gives a
 colour term of -0.32, somewhat lower than 
 that obtained by Stephens et al. (2000). 
At lower metallicities the coefficient decreases  to -0.2 at Z=0.0004.
The colour coefficients in this calibration are intermediate between
Stephens et al. and Cohn et al.'s.
The result of this test indicates that the metallicity is not 
the major source of the discrepancy between Stephens et al. (2000) and 
the synthetic photometry.  
We argue that an important source of discrepancy could be
reddening, in agreement with Lee et al. (2001).
Using their reddening law calculated for NICMOS bands they 
derived an average A(m$_{\rm 110}$)/E(B-V)=1.16 (for SGB and RGB stars), 
while Rieke and Lebofsky (1985) give A$_{J}$/E(B-V)=0.864. A star 
reddened by one magnitude in the m$_{\rm 110}$-m$_{\rm 160}$ colour will imply 
A(m$_{\rm 110}$)-A$_{\rm J}$=0.55 mag. 
It is evident that the colour terms coming 
from synthetic photometry are by far too small to account for the 
reddening effects.

We conclude that a correct calibration cannot be obtained if the 
reddening is not known in advance. A solution could be an iterative 
process deriving a first guess reddening and proceeding further 
with subtraction of the reddening and calibration of the dereddened data, 
adding the obtained final reddening to the calibrated data.

\begin{acknowledgements}
BB and EB acknowledge grants from CNPq and FAPESP.
SO acknowledges the Italian Ministero dell'Universit\`a e della Ricerca
Scientifica e Tecnologica (MURST) under the program
on 'Fasi iniziali di Evoluzione dell'Alone e del Bulge Galattico'
(Italy).
\end{acknowledgements}

%--------------------------------- References -------------

\end{document}